# Direct control of magnetic chirality in NdMn$_2$O$_5$ by external electric field

I.A. Zobkalo,[1] A.N. Matveeva,[1] A. Sazonov,[2] S.N. Barilo,[3] S.V. Shiryaev,[3] B. Pedersen,[4] V. Hutanu[2]

[1]*B.P Konstantinov Petersburg Nuclear Physics Institute, Kurchatov Institute, Gatchina, 188300, Russia;*
[2]*Institute of Crystallography, RWTH Aachen University and Jülich Centre for Neutron Science (JCNS) at Heinz Maier- Leibnitz Zentrum (MLZ), 85747 Garching, Germany;*
[3]*Scientific-Practical Materials Research Centre NAS of Belarus, Minsk, 220072, Belarus;*
[4]*Heinz Maier- Leibnitz Zentrum (MLZ), Technische Universität München, 85748 Garching, Germany.*

Detailed investigation of the incommensurate magnetic ordering in a single crystal of multiferroic NdMn$_2$O$_5$ has been performed using both non-polarized and polarized neutron diffraction techniques. Below $T_N \approx 30.5$ K magnetic Bragg reflections corresponding to the non-chiral type magnetic structure with propagation vector **k**$_1$ = (0.5 0 $k_{z1}$) occurs. Below about 27 K a new distorted magnetic modulation with a similar vector $k_{z2}$ occurs, which is attributed to the magnetization of the Nd$^{3+}$ ions by the Mn-sub-lattice. Strong temperature hysteresis in the occurrence of the incommensurate magnetic phases in NdMn$_2$O$_5$ was observed depending on the cooling or heating history of the sample. Below about 20 K the magnetic structure became of a chiral type. From spherical neutron polarimetry measurements, the resulting low-temperature magnetic structure $k_{z3}$ was approximated by the general elliptic helix. The parameters of the magnetic helix-like ellipticity and helical plane orientation in regard to the crystal structure were determined. A reorientation of the helix occurs at an intermediate temperature between 4 K and 18 K. A difference between the population of right- and left-handed chiral domains of about 0.2 was observed in the as-grown crystal when cooling without an external electric field. The magnetic chiral ratio can be changed by the application of an external electric field of a few kV/cm, revealing strong magnetoelectric coupling. A linear dependence of the magnetic chirality on the applied electric field in NdMn$_2$O$_5$ was found. The results are discussed within the frame of the antisymmetric super-exchange model for Dzyaloshinsky-Moria interaction.

## I.  INTRODUCTION

Magnetic multiferroics have attracted increasing scientific interest in the last fifteen years. In this class of materials magnetic order induces the long-range polar order and vice versa, thus both order parameters are strongly coupled. A coupling between two ferroic orders can induce novel functional properties that do not exist in either state alone. This fact provides technological interest to such materials, in particular to their potential use in various areas of spintronics, but also in the context of the fundamental understanding of various microscopic mechanisms leading to the magneto-electric coupling.

Manganite oxides family $RMn_2O_5$ (R – rare-earth element) represents a prominent example of multiferroics with an extremely interesting and close relationship between magnetism and ferroelectricity [1-3], including a magnetically induced polarization flop in $TmMn_2O_5$ [4], a polarization flip in $TbMn_2O_5$, also induced by a magnetic field [1], or electrically induced switching of antiferromagnetic domains in $YMn_2O_5$ [5]. The temperature evolution of the magnetic structure of multiferroic $RMn_2O_5$ is rather complex, which, in a combination with a variety of magnetoelectric phenomena, makes these compounds extremely attractive from the standpoint of studying fundamental phenomena of interaction between different order parameters. The intriguing feature of the $RMn_2O_5$ family is that compounds with rare-earth ions having a radius smaller than that of $Nd^{3+}$ are multiferroics, while those with a larger radius are not. $NdMn_2O_5$ itself, considered to be non-multiferroic for a long time, is of particular interest as a "borderline" in this series. Especially, only recently observed emergence of the electrical polarization in the magnetically ordered phase of this compound and its much lower value compared to that in another $RMn_2O_5$ multiferroics [6], makes it an attractive candidate for the verification of different proposed mechanisms of multiferroicity in this type of materials.

Generally, all members of $RMn_2O_5$ family are considered to be isostructural within the orthorhombic centrosymmetric space group *Pbam* [7]. This structure consists of octahedra $Mn^{4+}O_6$ edge-shared along the *c* axis and pairs of tetragonal pyramids $Mn^{3+}O_5$ linked to two $Mn^{4+}O_6$ chains. However, there are some indications in the literature about the symmetry lowering in $RMn_2O_5$ from *Pbam* down to *Pb2₁m* polar group at low temperatures [8-10]. It was reported also that in fact the structure is non-centrosymmetric even at room temperature, and non-centrosymmetric monoclinic space group *Pm* ($\gamma = 90°$) was proposed rather than the centrosymmetric one *Pbam* [11]. In this structure, there are several nearly equivalent superexchange paths between manganese ions that lead to the essential exchange competition, which, in turn, results in non-collinear magnetic structures and complex magnetic phases [12]. Multiple magnetic ordered phases suggest sophisticated magnetic interaction schemes, so it is reasonable to suppose that several mechanisms could be responsible for the complicated magneto-electrical behavior in $RMn_2O_5$ compounds.

The magnetic order in $NdMn_2O_5$ single crystal was studied recently by the neutron scattering [6, 13, 14]. The magnetic ordering in this compound was found to have features different from those known for other $RMn_2O_5$ multiferroics with smaller ion radii. For $NdMn_2O_5$ the onset of the long-range magnetic order with incommensurate wave vector $\boldsymbol{k} = (0.5\ 0\ \sim 0.36)$ is observed at $T_N$

≈ 30 K, while for other RMn$_2$O$_5$ multiferroics the transition to magnetically ordered phase takes place at temperatures higher than 40 K. Below 30 K two magnetic phases were reported, their propagation vectors being close to each other: ***k$_1$*** = (0.5 0 $k_{z1}$), ***k$_2$*** = (0.5 0 $k_{z2}$). In Refs [6, 14] it is shown that two propagation vectors merge together at temperature below T = 15 K, while in [13] we show that $k_{z1}$, $k_{z2}$ components remain separated and increase monotonously with the temperature decrease down to 25 K, then remain constant with values $k_{z1}$ = 0.384(2), $k_{z2}$ = 0.395(2). Two additional transitions were observed at ~ 20 K and at ~ 5 K. The latter is connected with Nd$^{3+}$ subsystem alignment with ***k*** $_{Nd}$ = (0.5 0 0) [14]. The transition at 20 K, related to the ordering of Mn magnetic moments, is of another nature and is of particular interest, since at 20 K the onset of electric polarization in NdMn$_2$O$_5$ single crystal was observed [6]. The most significant issue in the studies of RMn$_2$O$_5$ is the understanding of the microscopic mechanism responsible for the spin-driven ferroelectricity in these compounds. The model of inverse Dzyaloshinsky-Moriya interaction (DMI) works well for magnetic multiferroics in the RMnO$_3$ family, like TbMnO$_3$, DyMnO$_3$. In this model, DMI favors the displacement of O$^{2-}$ ions, which enhances the DMI exchange interaction between non-collinearly ordered spins and breaks the inversion symmetry [15, 16]. It is important to note that in RMn$_2$O$_5$ multiferroics ferroelectric polarization was enhanced in the commensurate magnetic phases with magnetic moments close to collinear alignment [2, 17]. This provides considerable difficulty in the explanation of the origin of ferroelectric polarization by the inverse DM model. In connection with this, the exchange striction mechanism is used to explain the origin of ferroelectricity [10, 12]. This model refers to collinear spin orders, where the symmetric exchange striction causes two nearest ions with parallel spins to move closer to each other, while those with antiparallel spins to move farther. The application of this model to RMn$_2$O$_5$ seems to be reasonable since the antisymmetric exchange is associated with a rather weak spin-orbit coupling. However, from another point of view, the competition of superexchange interactions leads to an exchange frustrated structure [3], and in such a case, a weak antisymmetric interaction can have a significant impact on the magnetic properties and subsequently on the electric properties of such a system. Apparently, this magnetic frustration is the reason why a considerable number of experimental facts indicate that a significant contribution to a rich set of magnetic phenomena in RMn$_2$O$_5$ is arising due to DMI [18, 19]. There are a number of arguments also in favor of multiple microscopic mechanisms of magnetically induced ferroelectricity in RMn$_2$O$_5$ compounds. NdMn$_2$O$_5$ is an ideal compound for such a study as it belongs to the same multiferroic family on the one hand, but do not show a commensurate magnetic order on the other. In non-collinear incommensurate magnetic structures of the DMI origin, the direction of spin rotation is fixed by DMI, while in centrosymmetric systems the magnetic energy is independent of the sense of rotation [20]. In order to clarify the role of DMI in formation of magnetic and magneto-electric properties in $RMn_2O_5$ we performed investigations of magnetic chirality and its evolution with the temperature and applied electric field in $NdMn_2O_5$. For that we employed the polarized neutrons diffraction. It provides great sensitivity to the direction of the spin rotation in the magnetic structure and allows obtaining information inaccessible by other experimental techniques.

## II. EXPERIMENTAL DETAILS

Single-crystals of $NdMn_2O_5$ were grown by the flux-melt method described in [21]. The crystal thoroughly characterized in the previous neutron diffraction investigation [13] was used for the actual experiments. The approximate dimensions of the used sample are 3x4x5 mm$^3$ with the longest dimension along the c-axis.

The measurements for the crystal structure refinement were performed at the diffractometer RESI at FRM-II reactor (Heinz Maier-Leibnitz Zentrum, Garching, Germany, MLZ) [22] using neutron wavelength of 1.039 Å. Data collection was made with the image plate position-sensitive 2D detector MAR345. The 2D data reduction was performed using the EVAL-14 suite [23].

Magnetic neutron studies were performed at diffractometer POLI at MLZ [24, 25], which permits measurements both in unpolarized and polarized diffraction modes using the same neutron wavelength. Non-polarized diffraction using variably double-focused Si (311) monochromator and wavelength of 1.15 Å, resulting in high-intensity neutron flux of more than $10^7$ n(cm$^2$s)$^{-1}$ with high resolution, was employed in order to study the thermal evolution of the magnetic Bragg reflections in $NdMn_2O_5$.

For the polarized neutron diffraction, we used the technique of Spherical Neutron Polarimetry (SNP). It is implemented on POLI using the third generation zero-field Cryogenic Polarization Analysis Device CRYOPAD [26]. Both neutron polarization and analysis are performed using $^3$He spin filters. Such configuration is very efficient for short-wavelength neutrons from the point of view of a good resolution and a high flux of the polarized neutrons on the sample position, superior to the previous investigations. At the same time, because the polarizing (analyzing) efficiency of the filters relaxes (decreases) with time, that represents a main drawback of the technique. The corrections should be done to the data, which may lead to some decrease in the statistical accuracy of the measurement. The additional corrections were performed according to calculations described in Ref. [27]. Standard SNP axis convention was applied for the polarization analysis: *x*-direction coincides with the scattering vector *q*, *y*-direction lies in the scattering plane and is perpendicular to *x* and *z*-axis is oriented vertically, thus forming a right-handed Cartesian coordinate system.

During the experiment with polarized neutrons, the crystal *b*-axis was oriented vertically, *a* and *c* axes were laying in horizontal scattering plane. In this way, the magnetic satellites of type $\boldsymbol{k}$ = ($k_x$ 0 $k_z$) can be reached. For the measurements with the electric field, the crystal was mounted on a sample rod of the cryostat, between the electrodes built from aluminum plates as shown in Fig. 1. The electric field was applied in a vertical direction along the *b*-axis. The sample was placed into a standard FRM-II type top-loading cryostat. The essentials of measurements with high electric fields in cryogenic environments consist in fine-tuning the pressure of helium exchange gas continuously on a thin line, where it is possible to apply several kV/mm without an electric breakdown while maintaining the necessary temperature control of the sample. Prior to the experiment, a reliable setup for in-situ pressure control and regulation within the cryostat has been developed and calibrated. In this way, an electric voltage up to 5 kV (corresponding to max. 12.5 kV/cm electric field in the sample) could be applied during the experiment without an electric breakdown. The power supply device FUG HCB 20M-1000 has been used for high voltage generation and high precision digital multimeter FLUKE 8846A for the current monitoring in order to avoid the breakdowns.

### III. EXPERIMENTAL RESULTS

#### A. Crystal structure refinement

For the crystal structure refinement, the datasets were collected at three temperatures – 300 K, 24 K, 6 K. The observed nuclear reflections were successfully indexed in *Pbam* space group and we did not observe superstructure reflections that should indicate that the crystal symmetry is lower than *Pbam*. But we also can suppose that this could be connected with the extremely low intensity of those reflections, as it was observed in [11]. The structural refinement was then performed with FullProff suite software [28]. The results obtained within *Pbam* space group are in good agreement with those reported previously at 300 K [6, 7] and are presented in Table I. The same datasets were then used for the refinement within *Pm* space group, but the enhancement of the fit quality was negligible by the significantly increasing number of fitting parameters , which does not permit us to make the definite decision in favor of lower symmetry group.

#### B. Temperature evolution of the magnetic order by non-polarized neutron diffraction

In order to identify the magnetic Bragg reflections from different magnetic phases for the polarized neutron investigation on POLI, we first performed a non-polarized temperature-dependent neutron diffraction study using the same sample, instrument settings, and beam parameters as for the subsequent polarized experiment.

The onset of magnetic ordering was observed at about $T_N \sim 30.5$ K in accordance with previous studies [6, 13, 14]. Below this temperature the magnetic satellites corresponding to the incommensurate propagation vectors $\boldsymbol{k}_1 = (0.5\ 0\ k_z)$, $\boldsymbol{k}_2 = -\boldsymbol{k}_1 = (0.5\ 0\ -k_z)$, $k_z = 0.361(4)$ appear (Fig. 2). If fitted with one Gaussian peak, the value of $k_z$ increases smoothly from 0.361(1) to 0.3967(1) with the temperature decrease from 30.5 K down to ~10 K (Fig. 3a). Below that lock-in temperature the propagation vector component $k_z$ of 0.3967(1) does not change any more down to 4 K. When heated, the lock-in value for $k_z$ remains constant up to ~ 20 K showing a significant temperature hysteresis and decreases smoothly with steeper slope toward higher temperatures, reaching almost the same values as by cooling at about 30 K (Fig. 3a). Hysteretic behavior between cooling and heating in the same temperature region has been reported for the dielectric permittivity in NdMn$_2$O$_5$ powders [6], which combined with our data suggests a possible coupling between magnetic and electric properties. Fig. 3b shows the temperature evolution of the integrated intensity of the observed magnetic satellite $(2\ 0\ 0)^{-k2}$ fitted by one Gaussian peak on cooling and heating. Slight temperature hysteresis is observable also for the intensity, however in another temperature region comparing to the thermal evolution of $k_z$ value. The hysteresis is observed between 10 K and 21 K. Above that latter temperature, the change in peak intensity is exactly the same when cooled or heated. A pronounced change in the slope of the thermal evolution of the magnetic peak intensity is observed at about the same temperature ~ 21 K. This temperature is of special interest as it is the temperature, where the emergence of the electric polarization in NdMn$_2$O$_5$ is reported [6]. Similar anomaly in the integrated intensities at ~ 20 K of other magnetic satellites e.g. $(0\ 1\ 0)^{+k1}$, $(2\ 1\ 0)^{-k2}$, $(0\ 0\ 0)^{+k1}$, $(0\ 1\ 2)^{+k2}$ was observed also in the previous studies [6, 13]. In ref. [6] three different types of the magnetic reflections, in regard to the thermal evolution, were reported: those occurring at about 30 K and monotonously increasing down to the lower temperatures e.g. $(0\ 1\ 0)^{+k1}$. Others, like those mentioned above, also occurred at about 30 K, slowly increased down to about 20 K increasing sharply at lower temperatures. And the third type, which only appears below 20 K and then continuously increases with the temperature lowering e.g. $(0\ 0\ 1)^{+k2}$. Such complex behavior has been interpreted as the presence of different incommensurate magnetic phases with slightly different magnetic propagation vectors $k_{z1}$ and $k_{z2}$, which may coexist in certain temperature regions [6, 13, 14]. The authors of [6, 14] based on the measurements of neutron powder diffraction and tiny single crystals proposed a physical picture where between 30 and 15 K there is an incommensurate phase called ICM1 with the coexistence of two slightly different $k_{z1}$ and $k_{z2}$ incommensurate modulations and below 15 K those modulations merge into one structure with $k_{z2}$, called ICM2. The authors of [13] based on the single crystal diffraction suggest the coexistence of the of $k_{z1}$ and $k_{z2}$ incommensurate modulations for all temperatures between 30 - 2 K, $k_{z1}$ being a dominant magnetic phase in the region 30-20 K and $k_{z2}$ - the main magnetic phase at temperatures below ~ 21 K. The presence of multiple phases with close ***k*** vectors should be reflected also in our data. It is worth mentioning that hot-neutron single crystal diffractometer is not the best instrument to study the tiny splitting between two magnetic satellites in low-*q* region, nevertheless optimizing the detector slits-width and monochromator focusing, a certain optimization in the resolution may be reached. Fig. 4 shows the temperature dependence of the width of the magnetic satellite $(2\ 0\ 0)^{-k2}$ resulting from the fit with one Gaussian peak when cooled or heated. The resolution limit of the instrument in this setup was of about 0.012 r.l.u. Two main features can be derived from these results. A) Even at low temperature (< 10 K), where magnetic satellite does not change with temperature anymore, the absolute value of the peak width fwhm = 0.0168(2) is significantly higher than the resolution of the instrument. This observation is in a good agreement with results reported for powder peaks in [6] and may suggest either broadening of the peak due to some disorder in the magnetic structure or still the presence of the two very close $k_z$ structures, as suggested in [13], with the only difference that they should be of the similar magnitude. B) The peak broadening (up to twice the instrument resolution limit) is a clear indication of the presence of additional magnetic phases. Depending on the cooling or heating process, the broadening sets at different temperatures resembling somehow the hysteresis behavior observed for the peak positions and intensities shown in Fig. 3. Close inspection of the results of high-resolution powder diffraction from Fig. 6 in Ref. [6] reveals that indeed the primary magnetic phase $k_{z1}$ occurring at about 30 K does not persist down to the very low temperature, but rather down to ~22 K only; then an additional phase with close $k_{z2}$ appears at about 26 K and persists down to ~12 K. This phase can be easily identified by the presence of the 3*$k_{z2}$ modulation in that temperature region only, as it is seen in Fig. 6 in Ref. [6]. And finally, at about 21 K a third incommensurate phase with $k_{z3}$ appears, which does not change down to the lowest temperature studied. It is not clear whether results presented in the Fig. 6 of Ref. [6] are measured in the heating or cooling mode, however comparing this physical picture to our results a clear similarity to our results in the cooling mode could be identified. Thus, in certain temperature regions different magnetic phases may coexist, like $k_{z1}$ and $k_{z2}$ between 27 - 23 K and $k_{z2}$ and $k_{z3}$ between 26 - 11 K by cooling, leading to the observed peak broadenings. The observed hysteresis in this scenario may be related to the formation of the intermediate $k_{z2}$ structure. When cooled, Mn magnetic sub-lattice, which orders at higher temperatures, influences in some way (polarizes) the Nd magnetic sub-lattice. This results in the non-sinusoidal modulated structure $k_{z2}$ and another modulation $k_{z3}$. At temperatures close to 5 K, Nd orders spontaneously resulting in the independent magnetic order with commensurate propagation vector ***k***$_{Nd}$ = (1/2 0

0) in good agreement with previous results [6, 14]. Fig. 5 shows the observed temperature dependence of magnetic peak $(2\ 0\ 0)^{+k^{Nd}}$. Both Mn magnetic sublattice with $k_{z3}$ and Nd sublattice with $\boldsymbol{k}_{Nd}$ coexist at temperatures below 5 K. While heated, an inverse process of the magnetic influencing may be suggested: the Nd ordered subsystem "polarizes" the Mn sub-lattice in a way that $k_{z3}$ is maintained up to the higher temperatures and intermediate $k_{z2}$ phase occurs later and disappears quicker by heating toward $k_{z1}$ phase. This scenario well resembles the slow change of the $k_z$ when cooled and steep change when heated observed in Fig. 3. The same model is applicable to the consideration of the peak width evolution in Fig. 4, denoting some "flexibility" in the occurrence of the phase $k_{z2}$ depending on cooling or heating sample history. Two transition temperatures for the magnetic ordering of Nd in NdMn$_2$O$_5$ have been reported recently from the splitting of crystal fields levels [29]. According to this study, at 28 K ground-state Kramers Doublet splits into two sublevels. This splitting sharply increases below ~18 K and reaches a maximum at 4.5 K. Such behaviour resembles fairly well the thermal evolution of the magnetic structure observed within this study in the heating mode, denoting the influence of the Nd magnetic moment on the magnetic structure of the Mn sub-lattice.

### B. Magnetic structure in zero electric field by spherical neutron polarimetry

The SNP technique allows measuring the relationship between the polarization of the incident and scattered neutron beams. It allows distinguishing between polarization rotation occurring upon scattering from lattice of the ordered magnetic moments and depolarization due the presence of the magnetic domains. The experimental quantities obtained in SNP experiment for each Bragg reflection are the components $P_{ij}$ of the 3x3 polarization matrix $\boldsymbol{P}$ of the type: $\mathcal{P}_{ij} = \frac{I_{ij}^+ - I_{ij}^-}{I_{ij}^+ + I_{ij}^-}$, where the indices $i, j$ refer to one of the orthogonal directions $x, y, z$ of the initial and final (scattered from the sample) polarization correspondingly. The first subscript corresponds to the direction of the initial polarization vector, while the second is the direction of analysis. $I$ is the measured intensity with polarization parallel "+" and antiparallel "-" to the initial polarization direction $i$. In the general case in all elements $\mathcal{P}_{ij}$ the contributions from nuclear and magnetic scattering exists, some of them containing also nuclear-magnetic interference and chiral scattering terms. A more detailed description of the SNP technique can be found elsewhere [30].

Considering magnetic ordering in a crystal as a plane elliptical helix, in a general case one can describe it by the following expression:

$$\boldsymbol{M}(\boldsymbol{r}_n) = \boldsymbol{u}\mu_u \cos(\boldsymbol{r}_n \cdot \boldsymbol{k}) + \boldsymbol{v}\mu_v \sin(\boldsymbol{r}_n \cdot \boldsymbol{k}), \tag{1}$$

where $\boldsymbol{u}, \boldsymbol{v}$ – unit vectors, orthogonal each other, $\mu_u, \mu_v$ – corresponding amplitudes, $\boldsymbol{r}_n$ – a cell position, helix vector $\boldsymbol{m} = [\boldsymbol{u} \times \boldsymbol{v}]$, and $\boldsymbol{k}$ – helicoid propagation vector. The diffraction on the incommensurate magnetic structure produces pure magnetic reflections, free from nuclear contribution. In the SNP mode we measured only those elements of polarization matrix, which will give most valuable information in a case of incommensurate magnetic structure, namely elliptic parameters $\mathcal{P}_{yy}$ and $\mathcal{P}_{zz}$ and chiral parameters $\mathcal{P}_{yx}$ and $\mathcal{P}_{zx}$ as well as scaling parameter $\mathcal{P}_{xx}$, denoting total magnetic scattering, which should be equal to -1 for purely magnetic reflection [30]. In this case terms of interest can be expressed as [19]:

$$\mathcal{P}_{yy} = -\mathcal{P}_{zz} \sim \frac{\mu_u^2 \cos^2\beta - \mu_v^2}{\mu_u^2 \cos^2\beta + \mu_v^2} = \frac{R^2 \cos^2\beta - 1}{R^2 \cos^2\beta + 1} \quad (2)$$

$$\mathcal{P}_{yx} = \mathcal{P}_{zx} \sim \frac{2(1-2n_r)\mu_u \mu_v}{\mu_u^2 + \mu_v^2} = \frac{2(1-2n_r)R\cos\beta}{R^2 \cos^2\beta + 1} \quad (3)$$

In these expressions $\boldsymbol{v}$ is considered to be directed vertically, $\boldsymbol{u}$ - horizontally, $\beta$ is the angle between the helix vector $\boldsymbol{m}$ and the scattering vector $\boldsymbol{q}$ (Fig. 6), $R = \mu_u/\mu_v$ - ellipticity, $n_r$ – a portion of "right-hand" helices in crystal, $n_l = 1 - n_r$ – a portion of "left-hand" helices. Thus, the overall relative chirality for the $k_n$ structure could be denoted as: $Ck_n = n_l(k_n) - n_r(k_n)$.

It should be noted that because of low intensities of scattering from phase $k_{z1}$ and $k_{z2}$ above 22 K (Fig. 3b) the measurements of polarization matrix elements were performed in a heating mode only for satellites from phase $k_{z3}$ at 4 K and later at 18 and 20 K where according to Figs. 3a and 4 the influence of the $k_{z2}$ phase could be observed. The measured polarization matrix elements for two different magnetic Bragg satellites with $+\boldsymbol{k}$ and $-\boldsymbol{k}$ vectors and different angles $\beta$ (see Fig. 6) measured at 4 K, 18 K and 20 K are shown in Table II.

The presence of the non-zero chiral parameters $\mathcal{P}_{yx}$ and $\mathcal{P}_{zx}$ of the same sign is clear direct evidence that the underlying incommensurate magnetic structure at 4 K is of the chiral type (cycloidal or helical order) and not a flat spin-wave modulation as previously suggested [6]. The non-zero value of these components gives evidence about the non-equal population of the domains with the "right" and "left" chiral handedness. Using the data at 4 K from Table I the calculation gives an estimation for chirality in $k_{z3}$ phase $Ck_{z3} = 0.20(2)$. At 18 K one still observes some non-zero chiral terms, but statistics is not as good because of the low satellite intensity in comparison with the 4 K. However, the fact that both $\mathcal{P}_{yx}$ and $\mathcal{P}_{zx}$ components are of the same sign for both reflections and their sign is the same as for 4 K, may serve as an indication that at 18 K the chiral magnetic phase $k_{z3}$ is still persisting also keeping the same type of imbalance between two different chiral handedness. At 20 K the chiral components $\mathcal{P}_{yx}$ and $\mathcal{P}_{zx}$ could not be measured any more reliably. This cannot be explained solely by decreasing statistics, and is an intrinsic treasure of the changing magnetic structure.

The elliptic parameters $\mathcal{P}_{yy}$ and $\mathcal{P}_{zz}$ for both $(2\ 0\ 0)^{-k2}$ and $(0\ 0\ 0)^{+k1}$ reflections in Table I also change with temperature. The more pronounced effect takes place for $(0\ 0\ 0)^{+k1}$ satellite, where $\mathcal{P}_{yy}$ and $\mathcal{P}_{zz}$ change their signs between 4 K and 18 K. This is clear evidence that orientation of the magnetic moments in $k_{z3}$ structure changes significantly between those temperatures while heated. Comparing this result with the temperature evolution of the peak width from Fig. 4b, this reorientation may be ascribed to the emerging additional phase called $k_{z2}$ at an intermediate temperature. No orientation change is observable between 18 and 20 K by further temperature increase. Assuming helical type magnetic structure and using Eq. (2) it is possible to calculate resulting magnetic helix parameters like ellipticity $R$ and helix plane orientation. Obtained in this way the angle $\alpha$ between the magnetic helix plane and crystallographic $ab$ plane for the $k_{z3}$ magnetic structure is: 22.2(3.2)° at 4 K. At higher temperatures a helix changes its orientation: the major and minor elliptic axes are swept and the angle between the helix plane to the $ab$ plane is tilted to the $\alpha = -4.7(4)°$ at 18 K, and $\alpha = -10.4(1.2)°$ at 20 K respectively. The ellipticity of the magnetic structure $R$, denoting the ratio between major and minor axes of the helix in the plane, changes also with temperature. $R = 1$ would correspond to a perfectly circular helix. Deviations from 1 denote distortions (elongations) of the ellipse. Calculated for different temperatures $R$ values are: 0.77(13), 2.16(11), 3.06(50) for 4, 18, and 20 K respectively, showing clear shrinking (elongation) of the helix toward the transition temperature around 21 K, where the chiral character disappeared (as shown in the next chapter), and plane spin density wave modulation is formed. The schematic representation of such transformation from chiral helical structure to non-chiral collinear spin-wave one is shown in Fig. 7.

### D. Magnetic chirality under applied electric field by polarized neutron diffraction

Because of low intensities of magnetic satellites, for the study of electric field dependence on chiral scattering, in order to increase statistics, we applied the technique of polarized neutron diffraction without the analysis of scattered beam polarization. In this case the loss of neutrons due to the limited transmission of the analyzer is avoided, while chiral scattering can be unambiguously detected. The expression for the intensity of polarized neutrons, scattered on a magnetic satellite from helical structure [31], can be written in the simplified form as:

$$I_x^{\pm} \sim \boldsymbol{M}_\perp \boldsymbol{M}_\perp^* \mp iCk_n \left( \boldsymbol{M}_\perp \times \boldsymbol{M}_\perp^* \right)_x \tag{4}$$

where $\boldsymbol{M}_\perp$ – effective magnetic moment participating in scattering process, perpendicular to scattering vector $\boldsymbol{q}$:

$$\boldsymbol{M}_\perp = \boldsymbol{q} \times (\boldsymbol{M} \times \boldsymbol{q}) \tag{5}$$

and $I_x^{\pm}$ - intensity of scattered neutrons with initial neutron polarization along $x$-axis (+) or opposite to it (-) and $x$-axis is directed along the scattering vector $q$. In this way, total magnetic intensity $I_M = (I_x^+ + I_x^-)/2$ as well as pure chiral part of the magnetic scattering $I_{Ch} = (I_x^+ - I_x^-)/2$ can be obtained if $Ck_n \neq 0$. At the temperature 4 K the results for chirality obtained by this technique were compared with those from SNP for the consistency.

The temperature evolution of chiral scattering, measured on magnetic satellite $(2\ 0\ 0)^{-k2}$ without electric field is shown in Fig. 8a. It is easy to see that the chiral scattering appears below $T \approx 21$ K. This means that below $T_N \approx 30$ K in NdMn$_2$O$_5$ a non-chiral incommensurate magnetic ordering emerges, while at $T_{Ch} \approx 21$ K the transition to chiral structure takes place. The crossover observed in the temperature dependence of the integrated intensities of magnetic satellites reflect this transition and the proposed $k_{z3}$ magnetic structure is of chiral nature. The existence of the non-chiral spin wave modulation phase in the intermediate temperature region between paramagnetic and ferroelectric phases is observed also for type-II multiferroic compounds – members of perovskite manganites family RMnO$_3$ with R = Tb, Dy [32-34] as well as some doped RMn$_2$O$_5$ family representatives like YMn$^{4+}$(Mn$_{0.88}$Ga$_{0.12}$)$^{3+}$O$_5$ [18].

In order to study the effect of the electric field on the magnetic properties of NdMn$_2$O$_5$ we first applied +5 kV voltage along $b$ direction (which yields electric field +12.5 kV/cm) at a fixed low temperature of 4 K while the sample was cooled without electric field (ZFC). This action had no measurable effect on the magnetic satellites. Then the measurements in the field cooled (FC) mode were performed in the following way: crystal was warmed up to the temperature of 40 K, i.e. well above the transition temperature $T_N$ = 30.5 K, then the voltage was applied and the sample cooled down to 5 K under the applied electric field. The sample was warmed up back to 40 K under the same field, then the voltage value was changed. The following sequence of applied fields was used: +5 kV (+12.5 kV/cm), +2.5 kV (+6.25 kV/cm), 0 kV, -2.5 kV (-6.25 kV/cm), -5 kV (-12.5 kV/cm). The intensities $I_x^+$ and $I_x^-$ were measured for selected magnetic satellites while cooling/heating under field at temperatures of 25 K, 20 K and then with 2.5 K step by temperature lowering down to 5 K, and then while heating at the same temperature points. The results of such measurements for satellite $(2\ 0\ 0)^{-k2}$ are presented in Fig. 8b. The application of the electric field in FC mode does not change $T_{Ch}$ for the same cooling/heating mode, while the intensity of the chiral scattering changes considerably. Depending on the field polarity certain chiral domains will be favored or suppressed. The behavior seems to be symmetric in regard to the pre-existed imbalance of domains with two handedness in the sample at zero field. Some temperature hysteresis of few Kelvin for $T_{ch}$ around 20 K depending on the heating or cooling mode was observed for chiral scattering. One should point out that the application of electric field (FC mode) did not change elliptic terms within the accuracy of the measurements. In Fig. 9 the chirality $Ck_3$ is plotted as a function of the applied electric field. This pretty much linear behavior is observable for the both phases. This is different comparing to the situation observed in TbMn$_2$O$_5$ [19], where similar measurements reveal a kind of hysteretic behavior in the chirality vs. electric field phase diagram. This result denotes weaker pinning of the chiral domain walls to the crystal frame in NdMn$_2$O$_5$, than in the case of TbMn$_2$O$_5$.

## IV. DISCUSSION

The emergence of the chiral scattering was observed at the temperature $T_{Ch}$ which coincides with $T_{CE}$ where ferroelectric polarization starts to occur. This leads us to the suggestion that namely the chiral magnetic structure is the origin of the ferroelectric polarization in NdMn$_2$O$_5$. Here it is appropriate to mention that antisymmetric DMI determines the direction of rotation of the spirals in the chiral magnetic structures [20]. The possibility of the existence of non-zero integral DMI in RMn$_2$O$_5$ was discussed previously in YMn$_2$O$_5$ [18] and recently for TbMn$_2$O$_5$ [19]. The considerations use a qualitative model based on the antisymmetric super-exchange through an anion [34, 35] taking into account the crystal symmetry of the non-centrosymmetric space group *Pm* rather than *Pbam* [11]. Apparently, the argumentation made there for TbMn$_2$O$_5$ is applicable also for NdMn$_2$O$_5$, which we repeat briefly here.

The antisymmetric super-exchange through an anion can be expressed as [35, 36]:

$$V_{DM} = \boldsymbol{D}[\boldsymbol{S}_1 \times \boldsymbol{S}_2], \quad (5)$$

where $\boldsymbol{D} = d(\theta)[\boldsymbol{R}_1 \times \boldsymbol{R}_2]$ is Dzialoshinsky vector in this case. $\boldsymbol{R}_1$, $\boldsymbol{R}_2$ are unit vectors, connecting magnetic ions with anion. Their vector product fixes the direction of $\boldsymbol{D}$. The antisymmetric exchange parameter $d(\theta)$ depends in a rather complex manner on the local anisotropy of metal ion, on the particularities of metal-ligand bonds and its sign depends on the bond angle $\theta$ [33]. The sign of the antisymmetric parameter $d(\theta)$ changes at a critical angle $\theta_k$ which is characteristic for each particular interacting cation-anion group. Within monoclinic space group *Pm* [11] the four pairs of interacting manganese ions Mn$^{3+}$ - Mn$^{4+}$ will give non-zero total interacting vector $\boldsymbol{D}$ (Fig. 10). For these pairs the Heisenberg exchange interactions are usually considered as $J_3$ and $J_4$. By the analogy, we will designate antisymmetric interactions under consideration as $\boldsymbol{D}_3$ and $\boldsymbol{D}_4$. In the case of non-centrosymmetric space group *Pm* there will be four inequivalent interaction $\boldsymbol{D}_{31}$, $\boldsymbol{D}_{32}$ and $\boldsymbol{D}_{41}$, $\boldsymbol{D}_{42}$. The bond angles on the regarded exchange paths are close to 130°, which is a critical super-exchange value according to Goodenough-Kanamori-Andersen rules [37].

It was shown for RMn$_2$O$_5$ [38] that oxygen locations relative to those of Mn$^{3+}$ and Mn$^{4+}$ ions play an important role in the determination of the magnetic properties. In this situation, a tiny deviation in anion position could cause a sharp change of the magnetic interaction [38]. This change should refer not only to the symmetric exchange but to the antisymmetric one as well. The comparison of the exchange bonds between the manganese ions as shown in Fig. 10 for $NdMn_2O_5$ and $TbMn_2O_5$ can be done based on the *Pbam* crystal structure, since, to our best knowledge, there is no exact data on the oxygen positions in $NdMn_2O_5$ and $TbMn_2O_5$ within space group *Pm* available. According to data from Ref. [7] the Mn11 – O43 – Mn23 and Mn11 – O44 – Mn24 angles and bonds corresponding to $D_{41}$, $D_{42}$ are similar for the both compounds. Whereas, for the pairs Mn11 – O31 – Mn21 and Mn11 – O32 – Mn22 (defining $D_{31}$, $D_{32}$) they are different. In $NdMn_2O_5$ there is a noticeable displacement of the oxygen ions O31, O32 away from the Mn11 ($Mn^{4+}$) toward the ions Mn21 and Mn22 ($Mn^{3+}$) respectively, comparing to $TbMn_2O_5$. The bond length Mn1-O3 is 2.076 Å for $NdMn_2O_5$ and 2.021 Å for $TbMn_2O_5$ respectively [7]. In both cases the angles are very close to the critical value of 132.4-132.5°.

Comparing the positions for the oxygen ions, for space group *Pm* [11] with those for *Pbam* [7] one can see that oxygen ions involved in DM interaction under our consideration O31, O32, O43, O44 have the largest polar displacements. Being the most "mobile" in this structure, these ions (or some of them) could undergo displacements under the influence of the electric field. These displacements provoke the change in the sign of the antisymmetric exchange parameter $d(\theta)$ in some domains, thus favoring predominant formation of the domains with certain handedness of the helices.

According to the obtained results of the chirality evolution under the applied electrical field we can suppose that geometrical arrangement of the considered super-exchange in $NdMn_2O_5$ is closer to the critical one than that of $TbMn_2O_5$, and therefore $NdMn_2O_5$ demonstrates less rigid chirality-to-lattice coupling. Crystal structure of $RMn_2O_5$ provides multiple competing magnetic interactions, and correspondingly it is exchange frustrated, which yields the competing magnetic ground states. For rare earth ions with the ion radius lower than that of $Nd^{3+}$ ground state with commensurate structure, characterized by $k$ = (0.5 0 0.25) becomes favorable in some temperature region, which is not the case for $NdMn_2O_5$. At the same time the spiral spin structure is still preserved in this commensurate phase in such compounds [19, 39]. Therefore, it is reasonable to assume that even in the in commensurate magnetic phase DMI interaction continues to persist and manifests in a considerable way generating weak ferroelectric polarization. The latter may be enhanced by some another mechanism, such as e.g. exchange striction [12] or/and oxygen spin polarization [40].

The $NdMn_2O_5$ system demonstrates the ability to generate ordered magnetic phases with slightly different *k*-vectors depending on the temperature. Much the same situation associated with the coexistence of several magnetic phases with similar wave vectors was observed also in $DyMn_2O_5$ [41]. Such a state can be realized due to the presence of several competing magnetic states. It was shown in [41] that the choice of the ground state can be obtained by a relatively small magnetic field, which reflects the proximity in the energy of the competing ground states in $RMn_2O_5$. In the case of $NdMn_2O_5$, the observed coexistence of magnetic phases can be associated with the choice of the different ground states occurring due to subtle deviations in the local environment of the manganese subsystem. Such a discrete splitting of the crystal field levels of the 4f electrons in $Nd^{3+}$ ions in $NdMn_2O_5$ under the temperatures studied here has recently been experimentally demonstrated by the infrared spectroscopy [29].

## V. SUMMARY

Detailed investigation of the complex incommensurate magnetic ordering in large single crystals of $NdMn_2O_5$ has been performed using polarized and non-polarized neutron diffraction techniques. Below $T_N \sim 30.5$ K a transition to the incommensurate magnetic order (most probably plane spin density wave modulation) with propagation vectors $\boldsymbol{k}_1 = (0.5\ 0\ k_z)$, $\boldsymbol{k}_2 = -\boldsymbol{k}_1 = (0.5\ 0\ -k_z)$ with $k_{z1} = 0.361(4)$ takes place. A few degrees below, at $T \approx 27$ K an additional incommensurate magnetic phase with similar propagation vectors $k_{z2}$ occurs. This phase, however, shows strong deviation from sinusoidal modulation as confirmed by the observation of the $3*k_{z2}$ modulations [6]. Interestingly, just at the same temperature an independent study of the CF energy levels of $Nd^{3+}$ reports the splitting of the ground-state Kramers doublet two sublevels [29], demonstrating a kind of "magnetic polarisation" of the Nd ions by the Mn magnetic sub-lattice. Thus, distorted $k_{z2}$ modulation may be attributed to the magnetisation of Nd sublattice. This magnetic phase seems to disappear, when the new chiral magnetic phase called $k_{z3}$ at lower temperature is fully formed. Significant temperature hysteresis is observed in the formation of the $k_{z2}$ and $k_{z3}$ phases depending on the heating or cooling sample history. Results reported here are obtained using a larger crystal of high quality and modern instrumentation with improved resolution and neutron flux in comparison to the previous studies [6,13,14] reporting somehow contradictory results about thermal evolution of incommensurate order in $NdMn_2O_5$. Our new findings allow unifying those results, and concluding that system demonstrates the ability to generate magnetic ordered phases of different types at temperatures between 30 and 4 K. According to the SNP measurements, below $T_{Ch} \approx 21$ K magnetic structure of $NdMn_2O_5$ becomes chiral and this kind of magnetic order is associated with the occurrence of the weak ferroelectricity. The resulting at lower temperature $k_{z3}$ magnetic structure could be well approximated using general elliptic helix model. Characteristic parameters of the helix, like ellipticity (ratio between minor and major ellipse axis) and inclination of the helix plane in regard to the crystallographic *ab* plane could be calculated for different temperatures. It was shown that while heating between 4 and 18 K a reorientation transition of the helical order (interchange between the minor and major axis orientation and change in the sigh on the inclination angle) happens. This reorientation may be caused by the occurrence of the $k_{z2}$ phase attributed to the partial magnetic ordering of the Nd ions at this temperature range observed in recent work [29]. Below ~ 5 K a commensurate AFM order with the propagation vector (1/2 0 0) is formed on Nd subsystem, this coexists with incommensurate structure $k_{z3}$. It is worth noting that we did not perform explicit investigations of the magnetic ordering of Nd within this study. With the temperature increase toward $T_{Ch}$, magnetic ellipse on Mn becomes more stretched along *a*-axis. The difference in the population of the "right" and "left" handedness domains in the chiral magnetic phase was observed in as-grown crystal. This difference can be controlled by the external electric field in field cooled mode. The linear dependence of the magnetic chirality from the applied electric field strength was observed within the used field region of ±12.5 kV/cm. Thus, strong magneto-electric coupling for $NdMn_2O_5$ was demonstrated experimentally for the first time. The results are qualitatively discussed within the frame of antisymmetric DMI super-exchange. Concluding for a more general case that integral antisymmetric DM exchange may exist in $RMn_2O_5$ manganates for the structures with the symmetry lower than *Pbam*. Our experimental findings are in good agreement with the suggestion that the DMI mechanism can be responsible for the emergence of weak ferroelectricity in $RMn_2O_5$ multiferroics family where the weak ferroelectric phase is observed simultaneously with incommensurate magnetic chiral structure.

## ACKNOWLEDGMENTS

The authors are grateful to S.V. Gavrilov for technical assistance and to S.V. Maleyev for fruitful discussions. This work is supported by Russian Foundation for Basic Research grant No. 16-02-00545-a. The work is based on the results obtained on instrument POLI, operated by RWTH Aachen in cooperation with FZ Juelich (Juelich-Aachen Research Alliance JARA).

______________________________________________________________________________

**Table captions**

TABLE I. Crystal structure parameters obtained within *Pbam* space group.

TABLE II. Polarization matrix elements for magnetic satellites $(2\ 0\ 0)^{-k2}$ and $(0\ 0\ 0)^{+k1}$ measured at different temperatures and zero electric field.

**Table I**

| 300 K ($a = 7.4940$ Å, $b = 8.6071$ Å, $c = 5.6934$ Å) | | | |
|---|---|---|---|
| Atom | X | Y | Z |
| Nd | 0.1421(3) | 0.1732(4) | 0 |
| $Mn^{4+}$ | 0 | 0.5 | 0.2556(8) |
| $Mn^{3+}$ | 0.4114(7) | 0.3539(7) | 0.5 |
| O1 | 0 | 0 | 0.2717(5) |
| O2 | 0.1579(5) | 0.4482(6) | 0 |
| O3 | 0.1520(5) | 0.4366(5) | 0.5 |
| O4 | 0.4023(3) | 0.2070(4) | 0.2497(4) |
| 24 K ($a = 7.4871$ Å, $b = 8.6293$ Å, $c = 5.7083$ Å) | | | |
| Nd | 0.1424(6) | 0.1724(5) | 0 |
| $Mn^{4+}$ | 0 | 0.5 | 0.257(1) |
| $Mn^{3+}$ | 0.411(1) | 0.354(1) | 0.5 |
| O1 | 0 | 0 | 0.276(1) |
| O2 | 0.1588(8) | 0.4482(6) | 0 |
| O3 | 0.1514(8) | 0.4366(6) | 0.5 |
| O4 | 0.4040(5) | 0.2050(5) | 0.2485(7) |
| 6 K ($a = 7.5135$ Å, $b = 8.6328$ Å, $c = 5.7320$ Å) | | | |
| Nd | 0.1436(7) | 0.1722(6) | 0 |
| $Mn^{4+}$ | 0 | 0.5 | 0.255(1) |
| $Mn^{3+}$ | 0.413(1) | 0.355(1) | 0.5 |
| O1 | 0 | 0 | 0.277(1) |
| O2 | 0.161(1) | 0.4501(7) | 0 |
| O3 | 0.1507(9) | 0.4353(7) | 0.5 |
| O4 | 0.4038(6) | 0.2059(5) | 0.2483(9) |

**Table II**

| 4 K | (2 0 0)$^{-k2}$ | | | (0 0 0)$^{+k1}$ | | |
|---|---|---|---|---|---|---|
| | x | y | z | x | y | z |
| x | -1.041(14) | | | -0.98(4) | | |
| y | -0.19(5) | -0.316(13) | | -0.20(2) | -0.20(2) | |
| z | -0.08(4) | | 0.292(14) | -0.19(2) | | 0.19(2) |
| 18 K | (2 0 0)$^{-k2}$ | | | (0 0 0)$^{+k1}$ | | |
| | x | y | z | x | y | z |
| x | -0.98(7) | | | -0.98(10) | | |
| y | -0.05(7) | -0.580(22) | | -0.19(5) | 0.35(2) | |
| z | -0.09(6) | | 0.584(24) | -0.06(5) | | -0.36(2) |
| 20 K | (2 0 0)$^{-k2}$ | | | (0 0 0)$^{+k1}$ | | |
| | x | y | z | x | y | z |
| x | -1.036(23) | | | -0.98(7) | | |
| y | | -0.728(21) | | | | |
| z | | | 0.721(21) | | | -0.53(10) |

**Figure captions**

Figure 1: NdMn$_2$O$_5$ single crystal sample (1) between two Al electrodes; (2) – isolated high voltage contact and (3) – ground; on the sample rod of FRM II close cycle refrigerator able to control temperature between 3.8 - 325 K with the precision of better than 0.1 K.

Figure 2: Q-space mapping of incommensurate magnetic Bragg satellites with propagation vectors $\pm \mathbf{k} = (1/2\ 0\ \pm k_z)$ in NdMn$_2$O$_5$ at 5 K.

Figure 3: Temperature dependence of: a) the $k_z$ component of magnetic propagation vectors $\mathbf{k}_{1,2}$ and b) the integrated intensity of the magnetic reflection $(2\ 0\ 0)^{-k2}$. Both a) and b) are plotted based on the fitting of magnetic reflection with one Gaussian peak. Where not visible, the error bars are smaller than the symbols. Lines are guides for the eyes.

Figure 4: Temperature dependence of the full width at the half maximum (FWHM) for the magnetic Bragg reflection $(2\ 0\ 0)^{-k2}$ as obtained from the fit with one Gaussian peak, a) by cooling and b) by heating. Data points are presented by the black squares, the red lines are the smoothening function by 3 neighboring points applied to the data as a guide for the eyes. The black arrows indicate the temperature region where an "intermediate" magnetic phase with $k_{z2}$ is supposed to exist (see main text for more details).

Figure 5: Temperature dependence of the magnetic Bragg reflection $(2\ 0\ 0)^{+k^{Nd}}$ by heating with commensurate propagation vector $\mathbf{k}_{Nd} = (1/2\ 0\ 0)$, associated with spontaneous ordering of the Nd$^{3+}$ magnetic sub-lattice.

Figure 6: Scheme of the consideration of directions in reciprocal space.

Figure 7. Schematic representation of the transformation of an elliptical spiral into a transverse spin-wave modulation in NdMn$_2$O$_5$ with temperature increase: (a) almost circular helix with minor axis along *a* – corresponding to configuration at 4 K, (b) – transformed (elongated) helix with major axis along *a* at 18-20 K, and (c) – plane spin wave modulation presumably at $T > 21$ K.

Figure 8. a) Temperature evolution of chiral scattering $I_{Ch}$, measured on satellite $(2\ 0\ 0)^{-k2}$ in heating mode without electric field. b) The temperature evolution of chiral scattering $I_{Ch}$ on the same satellite as function of electric field on cooling and heating mode. Solid lines are guides for eyes.

Figure 9. Electric field dependence of magnetic chirality $C_{k3}$ measured at 4 K in FC mode.

Figure 10. Exchange paths for Mn$^{4+}$ - Mn$^{3+}$ pairs in RMn$_2$O$_5$ considering monoclinic space group *Pm*.

Figure 1.

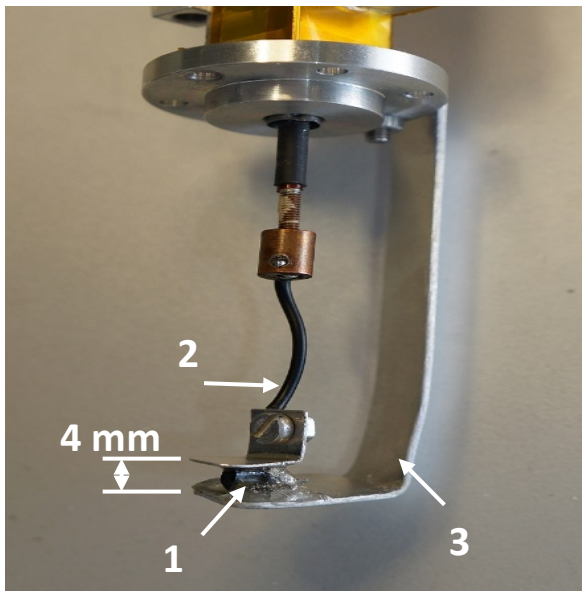

Figure 2.

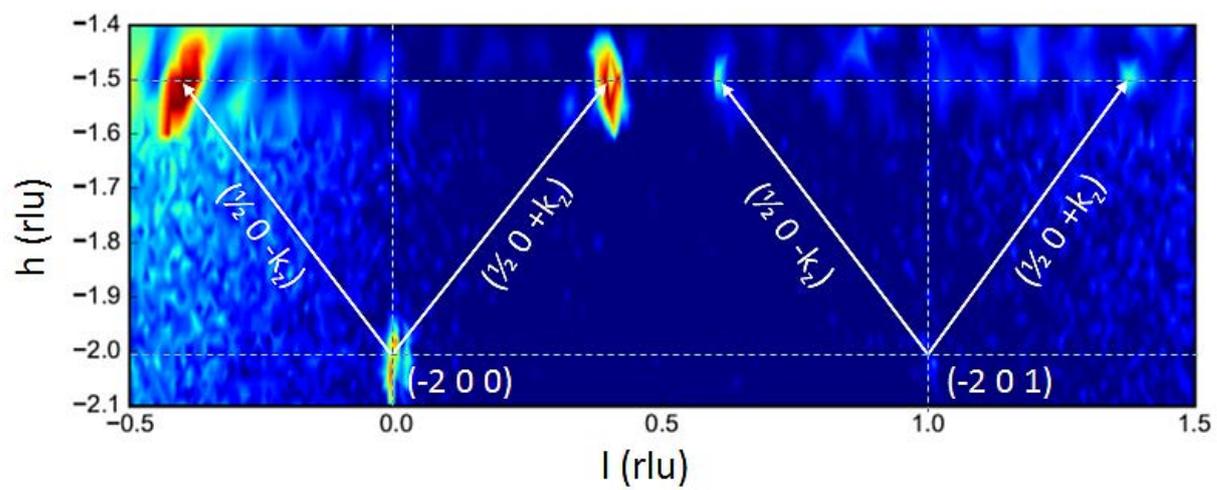

Figure 3.

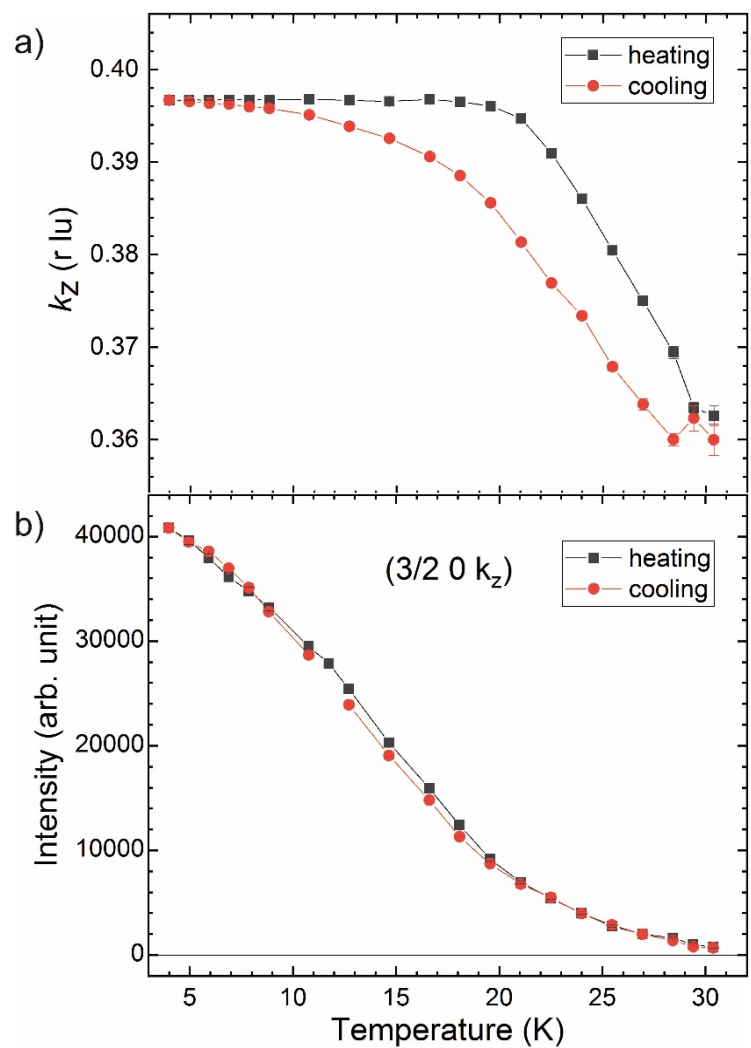

Figure 4.

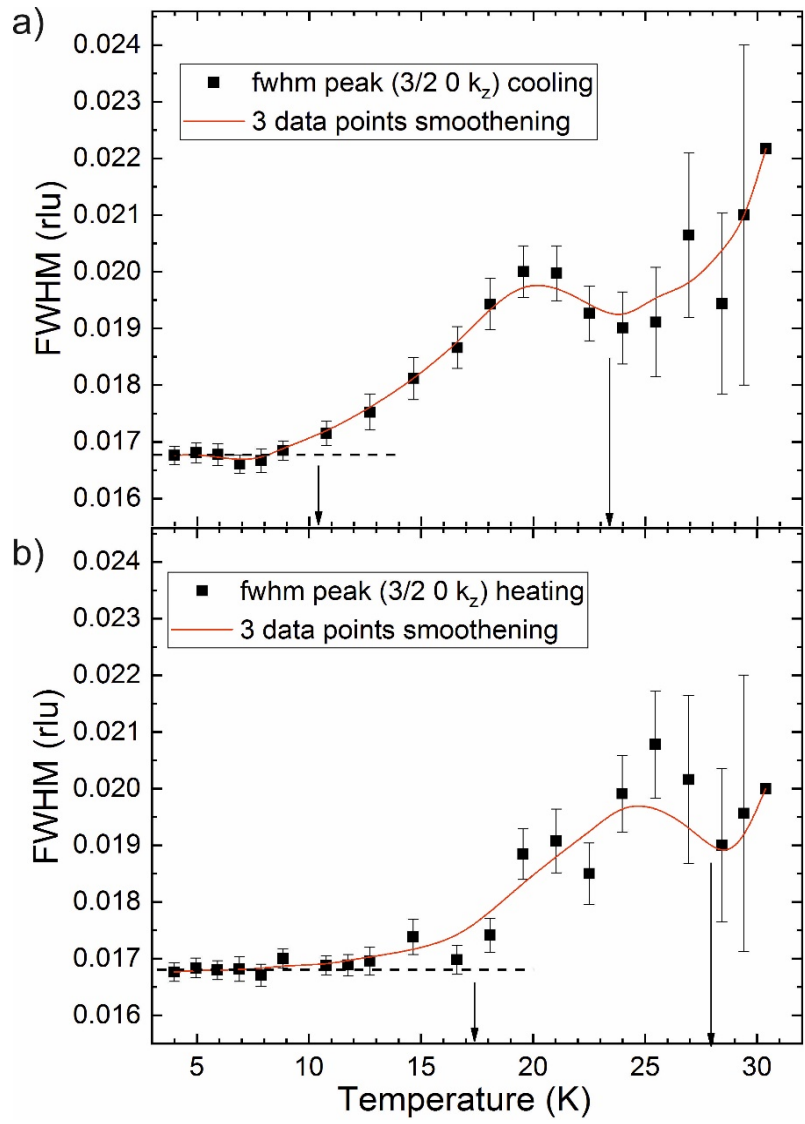

Figure 5.

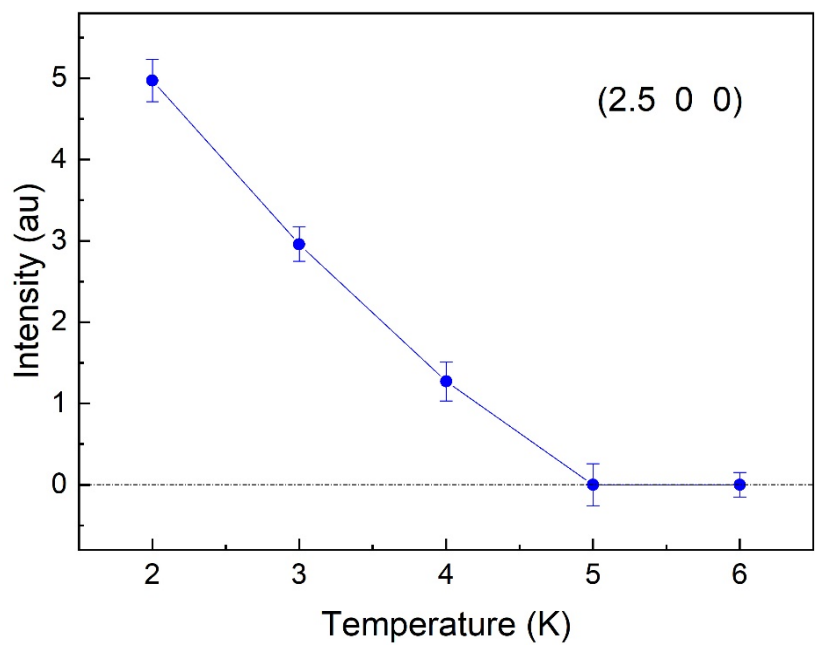

Figure 6.

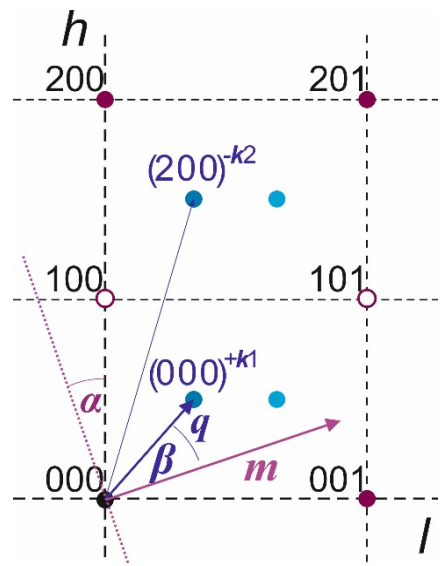

Figure 7.

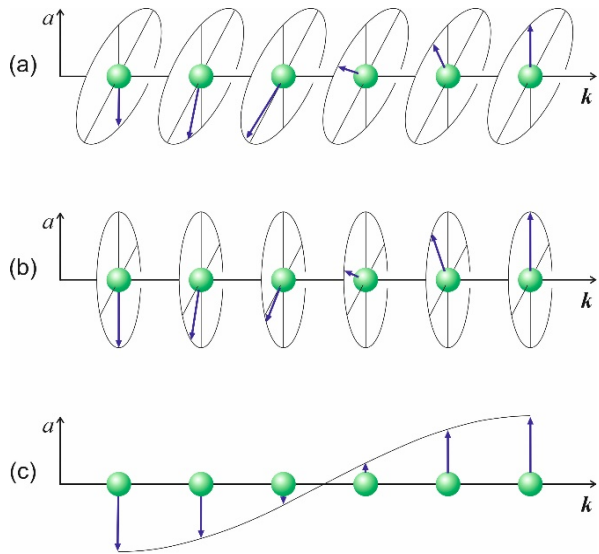

Figure 8.

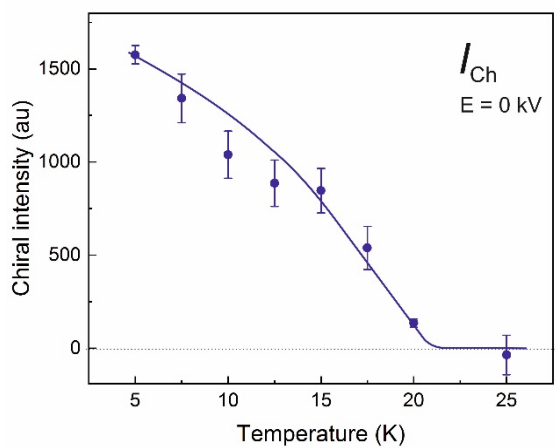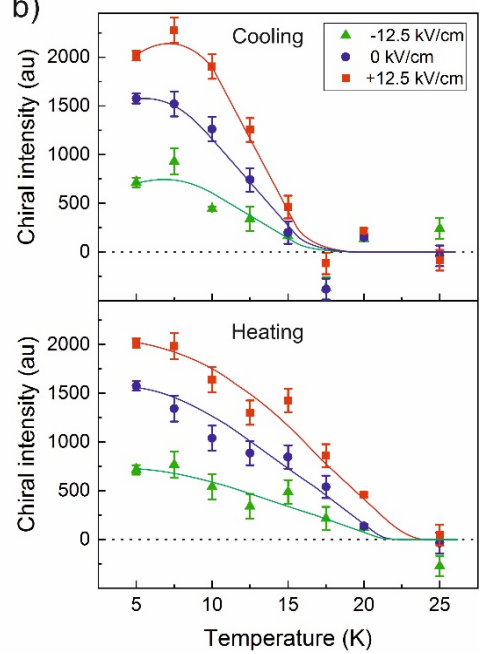

Figure 9.

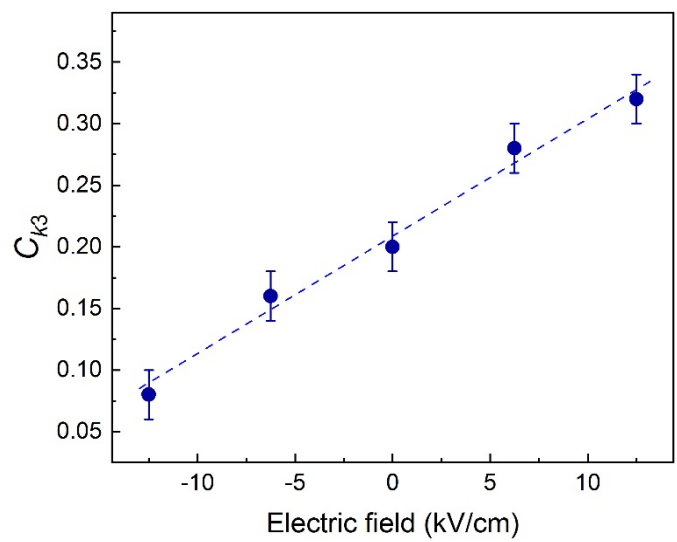

Figure 10.

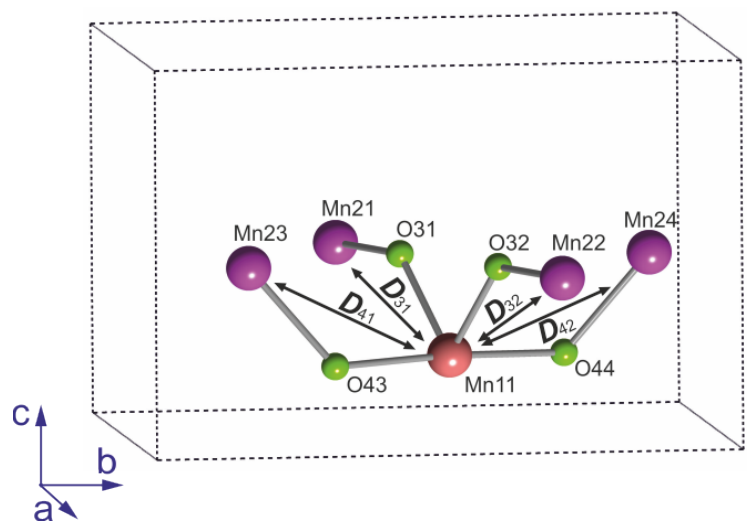